\documentclass[%
 reprint,
 amsmath,amssymb,
 aps,
floatfix,
]{revtex4-2}

\usepackage[utf8]{inputenc}
\usepackage{amssymb, amsthm, tabto, epigraph, mathtools, physics, dsfont, verbatim, slashed, mathrsfs}
\usepackage{graphicx, wrapfig}
\usepackage[shortlabels]{enumitem}
\usepackage{graphicx}
\usepackage{dcolumn}
\usepackage{hyperref}

\DeclareMathOperator{\Hom}{Hom}
\DeclareMathOperator{\bbZ}{\mathbb{Z}}
\DeclareMathOperator{\bbR}{\mathbb{R}}

\DeclareMathOperator{\Arf}{\textrm{Arf}}

\usepackage{tikz}
\usetikzlibrary{shapes,arrows,chains}
\usetikzlibrary{decorations.markings}
\usetikzlibrary{decorations.pathmorphing}
\tikzset{snake it/.style={decorate, decoration=snake}}

\begin{document}

\preprint{APS/123-QED}

\title{Two More Fermionic Minimal Models}

\author{Justin Kulp}
\email{jkulp@perimeterinstitute.ca}
\affiliation{%
 Perimeter Institute for Theoretical Physics, Waterloo, Ontario N2L 2Y5, Canada 
}%
\date{March 7, 2020}

\begin{abstract}
    In this short note, we comment on the existence of two more fermionic unitary minimal models not included in recent work by Hsieh, Nakayama, and Tachikawa. These theories are obtained by fermionizing the $\mathbb{Z}_2$ symmetry of the $m=11$ and $m=12$ exceptional unitary minimal models. Furthermore, these should be the only missing cases.
\end{abstract}

\maketitle

\section{INTRODUCTION}
In a recent paper by Hsieh, Nakayama, and Tachikawa, it was shown that there is a fermionic unitary minimal model for each $c = 1 - 6/(m(m+1))$. In particular, it is obtained by fermionizing the $\bbZ_2$ symmetry in the $A$ or $D$-type lattice models, and can be thought of as the fermionic partner to those two bosonic theories \cite{fermMinimalModels}.

The goal of this note is two-fold. First, we point out that the $m=11$ and $m=12$ exceptional unitary minimal models, also called $(A_{10},E_6)$ and $(E_6,A_{12})$, have a non-anomalous $\bbZ_2$ symmetry \cite{discreteSym}. From general considerations about fermionization, we conclude that there are an additional two fermionic minimal models which can be obtained from the generalized Jordan-Wigner transformation on these two models. Moreover, we use this note as an opportunity to illustrate some notation and simple ideas for upcoming work \cite{meDavide}.

\section{MINIMAL MODELS}
\subsection{Review and Classification}
Minimal models are (1+1)d CFTs whose Hilbert spaces are composed of a finite number of irreducible representations of the Virasoro algebra. A minimal model will generally have central charge
\begin{equation}
    c(p,p^\prime) = 1 - \frac{6(p-p^\prime)^2}{p p^\prime}\,,
\end{equation}
where $p$ and $p^\prime$ are positive co-prime integers and $p > p^\prime \geq 2$. The potential highest weights of a minimal model at $c(p,p^\prime)$ are given by the Kac formula
\begin{equation}
    h_{r,s} = \frac{(rp-sp^\prime)^2-(p-p^\prime)^2}{4p p^\prime}\,,
\end{equation}
where we use the symmetry $h_{r,s} = h_{p^\prime-r, p-s}$ of the Kac formula to produce a closed operator algebra of $(p-1)(p^\prime-1)/2$ distinct fields.

The modular invariants of minimal models are well known to have an ADE type classification \cite{cappelli1987conj, cappelli1987ade} (see also Tables 10.3 and 10.4 of \cite{dFMS}), which allows us to read off the highest weights/state spaces after identification with characters. More precisely, the modular invariants are in one-to-one correspondence with pairs of simply-laced Lie algebras. As a consequence of this classification, unless $p,p^\prime = 2,4$, there is always more than one modular-invariant minimal model at $c(p,p^\prime)$. That is, there are different operator algebras constructed from the same primaries and closed under OPE.

To obtain unitary minimal models, we specify to $(p,p^\prime) = (m+1, m)$ with $m\geq 2$, and can take $1 \leq s \leq r \leq m-1$. For example, at $m = 2$ we have the $c=0$ trivial CFT, at $m=3$ the $c=1/2$ critical Ising, and at $m=4$ the $c=7/10$ tricritical Ising. As promised, for $m = 5$ there are $A$-type and $D$-type modular invariants, which correspond to the tetracritical Ising and critical 3-state Potts model respectively. In general, for $m \geq 5$ there is always an $A$-type theory and a $D$-type theory, and at the special values $m=11,12,17,18,29,30$ there is a third $E$-type theory, corresponding to the Dynkin diagrams $E_6$, $E_7$, and $E_8$ for each consecutive pair. From here out we will only discuss unitary minimal models.

As mentioned in \cite{fermMinimalModels}, there is a temptation to call the $m=3$ theory a free massless Majorana fermion, and the $m=4$ theory the smallest $\mathcal{N} = 1$ supersymmetric minimal model. However, this is not strictly correct: the Ising model has only integral spin operators, and thus is bosonic, while the Majorana fermion has integral and half-integral spin operators, and thus is fermionic. It is similar for the $m=4$ case.


\subsection{Symmetries and \texorpdfstring{$\bbZ_2$}{Z2} Orbifold}
Symmetries of the operator algebra of a CFT are heavily restricted by conditions of unitarity and modular invariance. For example, $h=1$ operators are not unitary in a $c < 1$ CFT, so one can rule out continuous internal symmetries in minimal models \cite{confInvUnitary}. However, there may still be discrete symmetries.

In \cite{discreteSym}, the authors determine the maximal symmetry group of all unitary minimal models by studying the theories in the presence of twisted boundary conditions as in \cite{ZuberBCs, CardyBCs}. Said differently, they determine if there are twisted partition functions which can be consistently added to the theory. Summarizing, their findings are that: \textit{All unitary minimal models have maximal symmetry group $\bbZ_2$, except 6. The critical and tricritical 3-Potts model have non-commuting $\bbZ_2$ and $\bbZ_3$ symmetry which combine to an $S_3$ symmetry, and the 4 $E_7$ and $E_8$ minimal models have no symmetry.} We note that these symmetries are the same as the automorphisms of the associated Dynkin diagrams. 


As explained in \cite{fermMinimalModels}, a theory $T$ (put on $S^1 \times \bbR$ for concreteness) with non-anomalous $\bbZ_2$ symmetry can be coupled to a background $\bbZ_2$-connection. We may then consider the untwisted Hilbert space $\mathcal{H}_{\mathrm{Un.}}$, and the twisted Hilbert space $\mathcal{H}_{\mathrm{Tw.}}$, depending on whether or not the background $\bbZ_2$ is trivial, i.e. if states have holonomy around $S^1$. Moreover, these Hilbert spaces may be further decomposed into states that are even or odd under the $\bbZ_2$ symmetry
\begin{align}
    \mathcal{H}_{T,\mathrm{Un.}} &= \mathcal{H}_{T,\mathrm{Un.}}^+ \oplus \mathcal{H}_{T,\mathrm{Un.}}^-\\
    \mathcal{H}_{T,\mathrm{Tw.}} &= \mathcal{H}_{T,\mathrm{Tw.}}^+ \oplus \mathcal{H}_{T,\mathrm{Tw.}}^-\,.
\end{align}
Intuitively, the gauged theory $T//\bbZ_2$ consists of both untwisted and twisted Hilbert spaces because it is a sum over all background connections, but it only has those states which are gauge-invariant (i.e. have a $+$ superscript). Hence we have that
\begin{align}
    \mathcal{H}_{T//\bbZ_2,\mathrm{Un.}} &= \mathcal{H}_{T,\mathrm{Un.}}^+ \oplus \mathcal{H}_{T,\mathrm{Tw.}}^+\\
    \mathcal{H}_{T//\bbZ_2,\mathrm{Tw.}} &= \mathcal{H}_{T,\mathrm{Un.}}^- \oplus \mathcal{H}_{T,\mathrm{Tw.}}^-\,.
\end{align}
More generally, when $G \neq \bbZ_2$, the splitting ``even'' and ``odd'' would be promoted to the projectors $P_\chi = \frac{1}{\abs{G}} \sum_g \chi(g) g$ for characters $\chi \in \hat{G}$, and we recover the statements that $T//G$ has an emergent $\hat{G}$ symmetry, and that $T//G//\hat{G} = T$.

For the unitary minimal models, it is (literally) a textbook result (see 10.7 of \cite{dFMS}) that the $\bbZ_2$ symmetry of any of the A-type theories can be gauged to produce the D-type theory and vice-versa \footnote{It is shown in \cite{cheng2020relative} that the unitary minimal models do not have an `t Hooft anomaly.}.    

\section{FERMIONIC MINIMAL MODELS}
\subsection{Fermionization}
Given a bosonic theory $T_b$ with $\bbZ_2$ symmetry, it is possible to fermionize it by a generalized Jordan-Wigner transformation, turning the $\bbZ_2$ symmetry into a $(-1)^F$ Grassmann parity. Concretely, the partition function is obtained by summing over the $\bbZ_2$ connection and coupling appropriately to a spin-structure $\rho$. In particular, on a genus $g$ surface $\Sigma$, we have
\begin{equation}
    Z_{T_f}[\rho] = \frac{1}{2^g}\sum_{\alpha\in H^1(\Sigma,\bbZ_2)} (-1)^{\Arf[\alpha+\rho]} Z_{T_b}[\alpha]\,.
    \label{eq:spinFromBos}
\end{equation}

The invertible phases that can be stacked with any fermionic theory are classified by $\Hom(\Omega_d^\text{Spin}(pt), U(1)) = \bbZ_2$ \cite{kapustinThorngren:cobordism}. The effective action for the non-trivial invertible phase is given by the low energy version of the Kitaev chain
\begin{equation}
	e^{iS[\rho]} =  (-1)^{\Arf[\rho]}\,.
\end{equation}
Stacking with this theory changes the relative sign of the even and odd partition functions.

\begin{figure}[t]
	\centering
	\begin{tikzpicture}[baseline={(current bounding box.center)}]
	\tikzstyle{vertex}=[circle,fill=black!25,minimum size=12pt,inner sep=2pt]
	\node[vertex] (T_f) at (-1.7,0) {$T_f$};
	\node[vertex] (T_BH) at ( 2,2) {$T_b$};
	\node[vertex] (T_BL) at (2,-2) {$T^\prime_b$};
	\node[above] at (-2.5,0.5) {$\times \Arf$};
	\draw [-] (T_f) -- (T_BH) node[midway,above,rotate=+30] {Bosonize/Fermionize};
	\draw [-] (T_f) -- (T_BL) node[midway,below, rotate=-30] {Bosonize/Fermionize};
	\draw [-] (T_BH) -- (T_BL) node[midway,right] {Gauge $\bbZ_2$};
	\draw[<-] (-2,0.23) arc (-60+90:240+90:0.5);
	\end{tikzpicture}
	\caption{Gauging the $(-1)^F$ symmetry of a spin theory $T_f$ produces a bosonic theory $T_b$ with $\bbZ_2$ symmetry. A different bosonic theory $T^\prime_b$ is produced if one first stacks with the $\Arf$ theory. These two theories are related by $\bbZ_2$ orbifold.}
	\label{fig:gaugeGraphZ2Spin}
\end{figure}
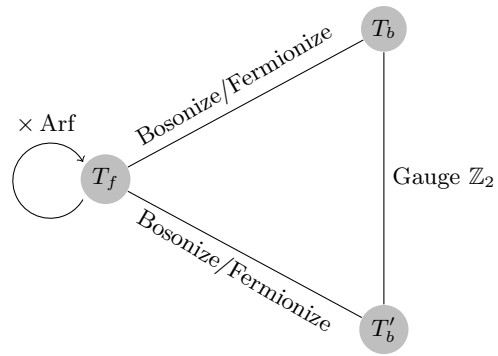

Conversely, given a fermionic theory $T_f$ (with $c-\bar{c} \in 8\bbZ$), it is possible to gauge the $(-1)^F$ symmetry and produce a bosonic theory $T_b$ with a $\bbZ_2$ symmetry in two distinct ways,
\begin{equation}
    Z_{T_b}[\alpha] = \frac{1}{2^g}\sum_{\rho}(-1)^{\Arf[\alpha+\rho]} Z_{T_f}[\rho]\,,
\end{equation}
or by stacking with $\Arf$ \footnote{It should be stressed that it does not make sense to think of one bosonization as being more fundamental. $T_f$ and $T_f \times \Arf$ are simply related by a trivial topological phase inherently linked to the Grassmann parity.} to get
\begin{equation}
    Z_{T_b^\prime}[\alpha] = \frac{1}{2^g}\sum_{\rho} (-1)^{\Arf[\alpha+\rho] +\Arf[\rho] } Z_{T_f}[\rho]\,.
\end{equation}
It is not hard to convince oneself, with the formulas above, that the two distinct bosonizations are related by gauging the emergent $\bbZ_2$'s, forming Figure \ref{fig:gaugeGraphZ2Spin}.

For a very concrete example, one might take the theory with $(-1)^F$ to be a free Majorana fermion $Z_{\textrm{Maj}}[\rho,M]$. Stacking with $\Arf$ amounts to switching the sign of the mass term, that is, $Z_{\textrm{Maj}}[\rho,-M] = (-1)^{\Arf[\rho]}Z_{\textrm{Maj}}[\rho,M]$. Bosonizing we attain the (1+1)d Ising model \cite{karchTongTurner:webOf2d}. Tuning to criticality, we see that the $M=0$ free Majorana fermion is the generalized Jordan-Wigner transformation of the critical Ising CFT. 

General points on fermionization and gauging are discussed further from a Hilbert-space and lattice-friendly point of view in \cite{fermMinimalModels}, and more from the point of view of partition functions in \cite{meDavide}.

\subsection{Fermionic Minimal Models}
We now briefly outline the work done by Hsieh, Nakayama, and Tachikawa in \cite{fermMinimalModels}. 

Combining the general theory above, the authors are effectively noting that the $A$ and $D$-type unitary minimal models are related by a $\bbZ_2$ orbifold. Then, as in Figure \ref{fig:gaugeGraphZ2Spin}, there must be a fermionic theory ``completing the triangle'' between the two bosonic theories.  In the case of the critical or tri-critical Ising, $T_b$ and $T_b^\prime$ are not distinct theories. This will be the case for the $E_6$ models, and provides a consistency check for us later, because clearly if a theory is self-dual under $\bbZ_2$ orbifold its fermionization must have vanishing $RR$-sector.

Due to the high degree of solubility of the unitary minimal models, the authors are able to explicitly tabulate the states in each of the theories and indicate which are twisted, untwisted, and even or odd (and similarly for their fermionizations). Furthermore, because the Jordan-Wigner transformation has a very concrete implementation when the theory is presented on a lattice, the authors use it to construct explicit lattice realizations of these fermionic minimal models from quantum spin-chains.

\section{The Exceptional \texorpdfstring{$E_6$}{E6} Cases}
We now come to the main point of this note, which is to point out the following: The $E_6$ exceptional minimal models at $m=11$ and $m=12$ also have a non-anomalous $\bbZ_2$ symmetry, and hence there are also fermionic minimal models associated with these.

Interestingly, the $E_6$ minimal models are self-dual under $\bbZ_2$ orbifold. This does make sense, there are no other unitary CFTs at the same central charge for them to transform into except possibly the $m=11,12$ $A$-type and $D$-type models, but clearly their $\bbZ_2$ symmetries have already ``been spent'' relating to one another.

We can be more explicit and list the partition functions of such theories by borrowing the expressions from \cite{discreteSym}. In particular, we use the twisted partition functions (equations (7.7) to (7.10) in \cite{discreteSym}) to do simple consistency checks on the statements above in the following subsections.

Lastly, now that all of the $\bbZ_2$ symmetries have been addressed, these exceptional $E_6$ cases, in tandem with the work in \cite{fermMinimalModels}, cover all possible fermionic unitary minimal models. Suppose otherwise, then there would be another fermionic model whose GSO projection gives a bosonic unitary minimal model with $\bbZ_2$ symmetry. As we have already determined the fermionizations of all such bosonic models, this additional fermionic model must not exist.

\subsection{\texorpdfstring{$m=11$}{m=11} Exceptional}
This theory is also known as the $(A_{10},E_6)$ unitary minimal model. The torus partition functions are
\begin{widetext}
    \begin{alignat}{2}
        Z_{(A_{10},E_6)}[0,0] &= \sum_{r=1,\mathrm{odd}}^{10} &&\abs{\chi_{(r,1)}+\chi_{(r,7)}}^2+\abs{\chi_{(r,4)}+\chi_{(r,8)}}^2+\abs{\chi_{(r,5)}+\chi_{(r,11)}}^2\\
        Z_{(A_{10},E_6)}[0,1] &= \sum_{r=1,\mathrm{odd}}^{10} &&\abs{\chi_{(r,1)}+\chi_{(r,7)}}^2-\abs{\chi_{(r,4)}+\chi_{(r,8)}}^2+\abs{\chi_{(r,5)}+\chi_{(r,11)}}^2\\
        Z_{(A_{10},E_6)}[1,0] &= \sum_{r=1,\mathrm{odd}}^{10} &&\abs{\chi_{(r,4)}+\chi_{(r,8)}}^2+\left\{(\chi_{(r,1)}+\chi_{(r,7)})^*(\chi_{(r,5)}+\chi_{(r,11)})+\mathrm{c.c.}\right\} \\
        Z_{(A_{10},E_6)}[1,1] &= \sum_{r=1,\mathrm{odd}}^{10} &&\abs{\chi_{(r,4)}+\chi_{(r,8)}}^2-\left\{(\chi_{(r,1)}+\chi_{(r,7)})^*(\chi_{(r,5)}+\chi_{(r,11)})+\mathrm{c.c.}\right\}
    \end{alignat}
\end{widetext}
It is easy to see that the spins coming from the untwisted Hilbert spaces are bosonic. With some computational work, one may verify that they take values $0, \pm1, \pm2, \dots, \pm 10, \pm 13, \pm 16, \pm 19$. 

One can also verify the partition function is invariant under modular $S$ and $T$ transformations, and that the twisted partition functions transform into one another under modular $S$ and $T$. It's also not hard to check that a $\bbZ_2$ orbifold returns the original partition function, or more generally that
\begin{equation}
    Z_{(A_{10},E_6)}[\alpha_1,\alpha_2] = \frac{1}{2}\sum_{\beta_1,\beta_2} (-1)^{\alpha_1 \beta_2 - \beta_1 \alpha_2} Z_{(A_{10},E_6)}[\beta_1,\beta_2]\,.
\end{equation}

The fermionic partition functions can be obtained from equation (\ref{eq:spinFromBos}) and are shown in equations (\ref{eq:m11fermions1})-(\ref{eq:m11fermions4}). As promised, the $RR$ sector, aka the periodic-bosonic sector, is vanishing. We also note that the fermionic sectors have both integral and half-integral operators, the spins from the untwisted NS-sector partition function are recorded in equation (\ref{eq:fermionicSpins}).
\begin{widetext}
    \begin{alignat}{2}
        Z_{f,11}[0,0] &= \sum_{r=1,\mathrm{odd}}^{10} &&\abs{\chi_{(r,1)}+\chi_{(r,7)}}^2+\abs{\chi_{(r,5)}+\chi_{(r,11)}}^2+\left\{(\chi_{(r,1)}+\chi_{(r,7)})^*(\chi_{(r,5)}+\chi_{(r,11)})+\mathrm{c.c.}\right\} \label{eq:m11fermions1}\\
        Z_{f,11}[0,1] &= \sum_{r=1,\mathrm{odd}}^{10} &&\abs{\chi_{(r,1)}+\chi_{(r,7)}}^2+\abs{\chi_{(r,5)}+\chi_{(r,11)}}^2-\left\{(\chi_{(r,1)}+\chi_{(r,7)})^*(\chi_{(r,5)}+\chi_{(r,11)})+\mathrm{c.c.}\right\}\\
        Z_{f,11}[1,0] &= \sum_{r=1,\mathrm{odd}}^{10} &&2\abs{\chi_{(r,4)}+\chi_{(r,8)}}^2 \\
        Z_{f,11}[1,1] &= \vphantom{\sum_{r=1,\mathrm{odd}}^{10}}0\label{eq:m11fermions4}
    \end{alignat}
    \begin{equation}
        \vphantom{\sum_{r=1,\mathrm{odd}}^{10}} 0,\pm\frac{1}{2},\pm\frac{2}{2},\dots,\pm\frac{10}{2},\pm\frac{13}{2},\dots,\pm\frac{17}{2},\pm\frac{20}{2},\pm\frac{21}{2},\pm\frac{25}{2},\pm\frac{26}{2},\pm\frac{29}{2},\pm\frac{32}{2},\pm\frac{35}{2},\pm\frac{38}{2},\pm\frac{45}{2}\label{eq:fermionicSpins}\,
    \end{equation}
\end{widetext}

\subsection{\texorpdfstring{$m=12$}{m=12} Exceptional}
This theory is also known as the $(E_6,A_{12})$ unitary minimal model. The torus partition functions are
\begin{widetext}
    \begin{alignat}{2}
        Z_{(E_6,A_{12})}[0,0] &= \sum_{s=1,\mathrm{odd}}^{12} &&\abs{\chi_{(1,s)}+\chi_{(7,s)}}^2+\abs{\chi_{(4,s)}+\chi_{(8,s)}}^2+\abs{\chi_{(5,s)}+\chi_{(11,s)}}^2\\
        Z_{(E_6,A_{12})}[0,1] &= \sum_{s=1,\mathrm{odd}}^{12} &&\abs{\chi_{(1,s)}+\chi_{(7,s)}}^2-\abs{\chi_{(4,s)}+\chi_{(8,s)}}^2+\abs{\chi_{(5,s)}+\chi_{(11,s)}}^2\\
        Z_{(E_6,A_{12})}[1,0] &= \sum_{s=1,\mathrm{odd}}^{12} && \abs{\chi_{(4,s)}+\chi_{(8,s)}}^2+\left\{(\chi_{(1,s)}+\chi_{(7,s)})^*(\chi_{(5,s)}+\chi_{(11,s)})+\mathrm{c.c.}\right\} \\
        Z_{(E_6,A_{12})}[1,1] &= \sum_{s=1,\mathrm{odd}}^{12} &&\abs{\chi_{(4,s)}+\chi_{(8,s)}}^2-\left\{(\chi_{(1,s)}+\chi_{(7,s)})^*(\chi_{(5,s)}+\chi_{(11,s)})+\mathrm{c.c.}\right\}
    \end{alignat}
\end{widetext}
As before, we can see the spins coming from the untwisted Hilbert spaces are bosonic, and that the modular $S$ and $T$ relationships between sectors is satisfied. The fermionic partition functions can be obtained from equation (\ref{eq:spinFromBos}) and are
\begin{widetext}
    \begin{alignat}{2}
        Z_{f,12}[0,0] &= \sum_{s=1,\mathrm{odd}}^{12} &&\abs{\chi_{(1,s)}+\chi_{(7,s)}}^2+\abs{\chi_{(5,s)}+\chi_{(11,s)}}^2+\left\{(\chi_{(1,s)}+\chi_{(7,s)})^*(\chi_{(5,s)}+\chi_{(11,s)})+\mathrm{c.c.}\right\}\\
        Z_{f,12}[0,1] &= \sum_{s=1,\mathrm{odd}}^{12} &&\abs{\chi_{(1,s)}+\chi_{(7,s)}}^2+\abs{\chi_{(5,s)}+\chi_{(11,s)}}^2-\left\{(\chi_{(1,s)}+\chi_{(7,s)})^*(\chi_{(5,s)}+\chi_{(11,s)})+\mathrm{c.c.}\right\}\\
        Z_{f,12}[1,0] &= \sum_{s=1,\mathrm{odd}}^{12} && 2\abs{\chi_{(4,s)}+\chi_{(8,s)}}^2 \\
        Z_{f,12}[1,1] &= \vphantom{\sum_{s=1,\mathrm{odd}}^{12}} 0
    \end{alignat}
\end{widetext}

\section*{Acknowledgement}
The author thanks Davide Gaiotto for pointing out that the exceptional minimal models should be considered and for comments on the draft. Research at Perimeter Institute is supported in part by the Government of Canada through the Department of Innovation, Science and Economic Development Canada and by the Province of Ontario through the Ministry of Colleges and Universities.

\bibliographystyle{apsrev4-2}
\bibliography{kulp_fermionic}

\end{document}